\documentclass[global]{svjour}
\usepackage{amsfonts}
\usepackage{amsmath}

\input{tcilatex}
\begin{document}

\title{Single-spin entanglement}
\author{G.B. Furman, \and V. M. Meerovich, \and V. L. Sokolovsky}
\institute{Physics Department, Ben Gurion University of the Negev%
%TCIMACRO{\TeXButton{email}{\email{gregoryf@bgu.ac.il}}}%
%BeginExpansion
\email{gregoryf@bgu.ac.il}%
%EndExpansion
}
\maketitle

\begin{abstract}
We show that the operators and the quadrupole and Zeeman Hamiltonians for a
spin $\frac{3}{2}$ can be represented in terms for a system of two coupling
fictitious spins $\frac{1}{2}$ using the Kronecker product of Pauli
matrices. Particularly, the quadrupole Hamiltonian which describes the
interaction of the nuclear quadrupole moment with an electric field gradient
is represented as the Hamiltonian of Ising model in a transverse selective
magnetic field. The Zeeman Hamiltonian, which describes interaction of the
nuclear spin with the external magnetic field, can be considered as the
Hamiltonian of the Heisenberg model in a selective magnetic field. The total
Hamiltonian can be interpreted as the Hamiltonian of $3D$ Heisenberg model
in an inhomogeneous magnetic field applied along the $x$-axis. The
representation of a single spin $\frac{3}{2}$ as two-spin $\frac{1}{2}$
system allows us to study entanglement in the spin system. One of the
features of the fictitious spin system is that, in both the pure and the
mixed states, the concurrence tends to $0.5$ with increase of applied
magnetic field. The representation of a spin $\frac{3}{2}$ as a system of
two coupling fictitious spins $\frac{1}{2}$ and possibility of formation of
the entangled states in this system open a way to the application of a
single spin $\frac{3}{2}$ in quantum computation.
\end{abstract}


\section{Introduction}

Entanglement is a quantum mechanical phenomenon in which the quantum systems
must be considered with reference to each other even they are spatially
separated \cite{1.NielsenMA
Chuang2000,2.BenentiG2007,3.AmicoL2007,4.HorodeckiR2009}. The paradoxical
behaviour of spin particular states, namely entangled state, has been
primarily pointed out by Einstein, Podolsky and Rosen \cite{4.EinsteinA}.
Today entanglement, as a long range quantum correlation between two or more
quantum systems,\ is considered as a well-established concept in modern
physics \cite{4a.K. Modi 2012,4b. Laflorencie2016,4c.G. Adesso2016}.

The unique properties of quantum entanglement and their important role in
modern physics have stimulated intensive investigation of various quantum
systems and search for measures and witnesses of entanglement \cite{4a.K.
Modi 2012,4b. Laflorencie2016,4c.G. Adesso2016}. In the last decade, the
quantum entanglement has received much attention in studies involving
quantum computing \cite{5. Bennett2000}, quantum communication \cite%
{6.Bennett}, and quantum metrology \cite{7. Roos,8. Cappellaro}. \ \ 

Since the first studies on quantum entanglement \cite{9.BellJ}, the models
consisting of spins $\frac{1}{2}$ have been intensively used as paradigms to
describe a wide range of many-body entangled systems. It is very convenient
to use spin models because each two-level system can be associated with a
spin $\frac{1}{2}$ placed in a {\small static} magnetic field \cite{10.Cohen}%
. Many phenomena and systems in quantum physics can be described applying
spin operator formalism because: (i) spin systems have a clear physical
picture, (ii) they can be controlled by a resonance radiofrequency field,
(iii) properties of the systems can be easily measured, for example, nuclear
magnetization by the nuclear magnetic resonance (NMR) technique, and (iv)
the spin $\frac{1}{2}$ systems can be relatively easily described
theoretically, since spin $\frac{1}{2}$ operators are defined by simple
operational transpositions and commutation rules.

To simplify the description of systems consisting of spins greater than $%
\frac{1}{2}$, various representations with projection operators \cite%
{10.Kessel1963,10a.Furman GB} have been employed. In particular, it was
shown that a nuclear spin $\frac{3}{2}$ can be represented by two spin
subsystems which are described using the blocks consisting of Pauli spin $%
\frac{1}{2}$ matrices $2\times 2$ \cite{12.Bloom,13.Das,14.G. W. Leppelmeier}%
. An alternative way of using fictitious spins $\frac{1}{2}$ for describing
a spin $\frac{3}{2}$ was proposed in \cite{10. M Goldman1988,10a. D.
Petit1988}. The multiple-quantum spin dynamics in systems with spin $1$ was
studied using a fictitious spin of $\frac{1}{2}$, forming the SU(3) group 
\cite{15.S.Vega}. However, all these attempts were limited to model where
the spin operators and Hamiltonians for spin greater than $\frac{1}{2}$ were
presented as a single fictitious spin \cite{15.S.Vega} or as a system
consisting of several non-coupled spins of $\frac{1}{2}$ \cite{14.G. W.
Leppelmeier}. Therefore, such an approach does not reflect realistic physics
situation and processes and allow a consistent analysis of quantum
phenomena, such as quantum entanglement, in the systems with a spin larger
than $1/2$.

The same approach was used to implement the quantum gate using two qubits
which are formed on the basis of a single quantum particle with spin $\frac{3%
}{2}$ \cite{16.A. R. Kessel,17.A. K. Khitrin} by applying technique of the
nuclear magnetic resonance (NMR) with quadrupole splitting \cite{16.A. R.
Kessel,17.A. K. Khitrin} and pure (without external magnetic fields) nuclear
quadrupole resonance \ (NQR) \ \cite{18.G.B.Furman,19.G.B.Furman,20.G. B.
Furman,21.Khitrin A}. This idea was confirmed in NMR experiments \cite{17.A.
K. Khitrin,21.Khitrin A}.

Because entanglement is an essential resource in current experimental
implementations for quantum information processing, it will be very useful
to study the conditions required to entangle qubits based on the states of a
single spin $\frac{3}{2}$. Recently the conditions for quantum states of
nuclei possessing quadrupolar moment to be entangled were studied with
presented a spin $\frac{3}{2}$ as a set of two qubits, which is isomorphic
to a qubit system consisting of two coupled spins $\frac{1}{2}${\Huge \ }in
our woks \cite{20.G. B. Furman,22a.G Furman2015}. It was obtained that
entanglement can be achieved by applying an external magnetic field to\ spin-%
$\frac{3}{2}$ nuclei in the electric field gradient (EFG) generated by
charges in their surroundings. However, in our previous works the analysis
of entanglement has not been based on explicit and confirmed representation
of a spin $\frac{3}{2}$ as a system of two spins $\frac{1}{2}.$ Such
consideration,\ as shown below, does not affect the results obtained for
quantum entanglements. \ One of the purposes of the present paper is
elimination of the disadvantage and to develop a way based on explicit
introduction of fictitious spins. This approach allows us to make clear the
interaction between the fictitious spins and their interaction with a
magnetic field, as well as possible their selective control.\textbf{\ }We
consider a nucleus with spin $\frac{3}{2}$ in an inhomogeneous electric
field and an external magnetic field and develop the description of states
of spin $\frac{3}{2}$ by identifying it with a system of two spins $\frac{1}{%
2}$ using the Kronecker product of the Pauli matrices. Thus, the particle
possessing spin $\frac{3}{2}$ can be considered as consisting of two parts
and we study the entanglement of these parts.

The paper structure is the following. In the next section we explicitly show
that the spin operators and the quadrupolar and Zeeman Hamiltonians for a
spin $\frac{3}{2}$ can be represented in the operator terms for a system of
two coupling fictitious spins $\frac{1}{2}$. In Section III, entanglement
for the system of two fictitious spins in the pure and mixed states is
studied. In the last section we conclude and discuss our results.

\section{Decomposition of the Hamiltonian for spin $\frac{3}{2}$ on the
basis of the Pauli matrices}

In a crystalline solid, the electric quadrupole moment , $eQ,$of a nucleus
possessing spin $\frac{3}{2}$, interacts with the gradient of the electric
field, $\frac{\partial ^{2}V}{\partial x_{i}\partial x_{j}}$ $\left(
x_{i},x_{j}=x,y,z\right) $ generated by or the surrounding electrons or
external charges of other nuclei. This interaction results in splitting of
the energy levels which are separated by distances proportional to the
quadrupole coupling constant $\frac{e^{2}Qq}{\hbar }$, where $eq=\frac{%
\partial ^{2}V}{\partial z^{2}}$, $V$ is the potential of the electric field
and $e$ is the proton charge \cite{13.Das,24.Abragam A}. In the principal
axis frame (PAF) with the $z$- and $x$-axes directed along the maximum and
minimum of the electric field gradient (EFG), respectively, $\left\vert
V_{zz}\right\vert \geq \left\vert V_{yy}\right\vert \geq \left\vert
V_{xx}\right\vert ,$ the EFG symmetric tensor is reduced to a diagonal form.
The quadrupolar Hamiltonian, $\mathcal{H}_{Q}$ , in the PAF takes the form
(we used units where $\hbar =1)$ \cite{13.Das,24.Abragam A}

\begin{equation}
H_{Q}=\omega _{Q}\left[ 3I_{z}^{2}-\vec{I}^{2}+\eta \left(
I_{x}^{2}-I_{y}^{2}\right) \right] ,  \tag{1}
\end{equation}%
with the quadrupole frequency $\omega _{Q}=\frac{e^{2}Qq}{4I\left(
2I-1\right) }$ and the asymmetry parameter\emph{\ }$\eta $ is defined as 
\begin{equation}
\eta =\frac{V_{yy}-V_{xx}}{V_{zz}},  \tag{2}
\end{equation}%
and may vary between $0$ and $1$. $I_{i}$ $\left( i=x,y,z\right) $ are the
projections of the spin angular momentum operator $\vec{I}$ on the $x$-,$y$%
-, and $z$-axes, respectively.

In the presence of an applied magnetic field $\vec{H}_{0}$ directed along
the $x$-axis of the PAF the Hamiltonian $\mathcal{H}$ of a nuclear spin can
be written:%
\begin{equation}
\mathcal{H=H}_{x}\mathcal{+H}_{Q}\text{,}  \tag{3}
\end{equation}%
where the Zeeman Hamiltonian $\mathcal{H}_{x}$ describes interaction of the
nuclear spin with the magnetic field

\begin{equation}
\mathcal{H}_{x}=-\gamma H_{0}I_{x},  \tag{4}
\end{equation}
where $\gamma$ is the nuclear gyromagnetic ratio.

Any operator of a single spin $\frac{1}{2}$ can be presented as a
combination of four $2\times 2$ Hermitian matrices \ 
\begin{equation}
\{\sigma _{0},\sigma _{x},\sigma _{y},\sigma _{z}\},  \tag{5}
\end{equation}%
\qquad\ \ \ \ where $\sigma _{0}=\hat{e}$ is the identity operator and $%
\sigma _{x}$, $\sigma _{y}$, $\sigma _{z}$ are the Pauli spin operators.
This set of matrices forms an orthonormal basis, meaning that the trace of
the product of any two different elements vanishes, while the trace of
square of each element equals $2$.

By analogy, any ${\small 4\times 4}$ matrix may be parametrized as a
superposition of 16 direct products of four Hermitian matrices (5). The set
of $16$ operators 
\begin{equation}
\sigma _{i}\otimes \sigma _{j}\text{ \ \ \ with }i,j=0,x,y,z.  \tag{6}
\end{equation}%
is orthonormal and complete for a spin $\frac{3}{2}$ in the same sense as a
set of four Hermitian matrices (5) for a single spin $\frac{1}{2}$: the
trace of the square of each operator equals $4$ \ and the trace of the
product of any two different matrices vanishes. As an example, three
projections $I_{i}$ $\left( i=x,y,z\right) $ of the spin angular momentum
operator $\vec{I}$ \ and the unit ${\small 4\times 4}$ matrix, $E,$ can be
presented by using set (6) in the following form\ 
\begin{align}
\bigskip I_{x}& =\frac{\sqrt{3}}{2}\sigma _{0}\otimes \sigma _{x}+\frac{1}{2}%
\sigma _{x}\otimes \sigma _{x}+\frac{1}{2}\sigma _{y}\otimes \sigma _{y}, 
\notag \\
I_{y}& =\frac{\sqrt{3}}{2}\sigma _{0}\otimes \sigma _{y}-\frac{1}{2}\sigma
_{x}\otimes \sigma _{y}+\frac{1}{2}\sigma _{y}\otimes \sigma _{x},  \notag \\
I_{z}& =\sigma _{z}\otimes \sigma _{0}+\frac{1}{2}\sigma _{0}\otimes \sigma
_{z},  \notag \\
E& =\sigma _{0}\otimes \sigma _{0}.  \tag{7}
\end{align}

Using (7), the quadrupolar Hamiltonian, $\mathcal{H}_{Q}$ and the Zeeman
Hamiltonian, $\mathcal{H}_{X}$ can be rewritten in terms of the Pauli
operators as%
\begin{equation}
\mathcal{H}_{Q}=\mathcal{\omega }_{Q}\left( 3\sigma _{z}\otimes \sigma _{z}+%
\frac{\sqrt{3}}{2}\eta \sigma _{x}\otimes \sigma _{0}\right) ,  \tag{8}
\end{equation}

\begin{equation}
\mathcal{H}_{x}=\mathcal{\omega }_{0}\left( \frac{\sqrt{3}}{2}\sigma
_{0}\otimes \sigma _{x}+\frac{1}{2}\sigma _{x}\otimes \sigma _{x}+\frac{1}{2}%
\sigma _{y}\otimes \sigma _{y}\right) .  \tag{9}
\end{equation}

Thus, the projections of angular momentum operator and the quadrupolar and
Zeeman Hamiltonians for a spin $\frac{3}{2}$ are represented in the operator
terms for a system of two coupling spins $\frac{1}{2}$.

The quadrupolar Hamiltonian (8) which describes the interaction of the
nuclear quadrupole moment with EFG represents the Hamiltonian of the Ising
model in a transverse selective magnetic field parallel to the $x$-axis. The
constant of the spin interaction is $3\mathcal{\omega}_{Q}$ and the strength
of the magnetic field is $\frac{\sqrt{3}}{2}\eta\mathcal{\omega}_{Q}$. At $%
\eta=0$ Hamiltonian (8) is reduced to the usual Ising Hamiltonian. Note,
that even in the case with $\eta=0$, Hamiltonian (8) cannot be represented
as a combination of the $\sigma_{z}\otimes\sigma_{0}$ and $%
\sigma_{0}\otimes\sigma_{z}$. Therefore in the general case a spin $\frac{3}{%
2}$ cannot be considered as a system of two non-coupled spins $\frac{1}{2}$.

Hamiltonian (9), which describes the interaction of the nuclear spin $\frac{3%
}{2}$ with the external magnetic field, can be considered as the Hamiltonian
of a $XY$ Heisenberg model in a selective magnetic field along the $x$-axis.
The total Hamiltonian (3) can be interpreted as the Hamiltonian of $3D$
Heisenberg model with an inhomogeneous magnetic field applied along the $x$%
-axis%
\begin{equation}
\mathcal{H}_{H}=J_{x}\sigma _{x}\otimes \sigma _{x}+J_{y}\sigma _{y}\otimes
\sigma _{y}+J_{z}\sigma _{z}\otimes \sigma _{z}+h_{01}\sigma _{x}\otimes
\sigma _{0}+h_{02}\sigma _{0}\otimes \sigma _{x},  \tag{10}
\end{equation}%
where $J_{x}=J_{y}=\frac{1}{2}\mathcal{\omega }_{0}$, $J_{z}=3\mathcal{%
\omega }_{Q}$, $h_{01}=$ $\frac{\sqrt{3}}{2}\eta \mathcal{\omega }_{Q}$ and $%
h_{02}=\frac{\sqrt{3}}{2}\mathcal{\omega }_{0}$. In contrast to the usual
Heisenberg model, the constants of the spin interaction $J_{x}$ and $J_{y}$
depend on the external field, while the constant $J_{z}$ is determined by
the quadrupole interaction. The magnetic field acting on the first
fictitious spin $\frac{1}{2}$ does not depend on applied field and is
determined only by the quadrupole interaction unlike the magnetic field
acting the\ second spin depends on the applied field only.

Therefore, Hamiltonian (3) for a spin $\frac{3}{2}$ is represented as the
Hamiltonian (10) which describes the system of two fictitious coupling spins 
$\frac{1}{2}$ in an inhomogeneous magnetic field. This reformulation allows
us interpreted the obtained results as a Hamiltonian of system of two
coupled fictitious spins $\frac{1}{2}$. The coupled constants depend on the
external field, which allows us to control the strength of the coupling
between the fictitious spins. The control of the strength of the interaction
between the spins makes it possible, on the one hand, to reduce the
execution time of quantum gates in the implementation of logic gates, and on
the other hand, to regulate the decoherence processes in such a spin system.

The energy levels of the spin system with Hamiltonian (8) are degenerate,
but the degeneracy is removed in the presence of an applied magnetic field
directed along the $x$-axis (9) and results in differences in the resonance
frequencies of the fictitious spins: $\Omega _{1}=\sqrt{3}\eta \omega _{Q}$%
\textbf{\ }for the first fictitious spin and $\Omega _{2}=$\ $\sqrt{3}\omega
_{0}$\ for the second fictitious spin. Therefore, an efficient way to
manipulate the fictitious spins is to irradiate the spin system with
selective radio frequency fields directed along the $z$-axis. The
Hamiltonian parts describing the acting on the first\ and the second spins
are $\mathcal{H}_{r.f}^{(1)}=\gamma H_{1}I_{z}\cos \Omega _{1}t$ and $%
\mathcal{H}_{r.f}^{(2)}=\gamma H_{2}I_{z}\cos \Omega _{2}t$, respectively.
Here $H_{1}$ and $H_{2}$ are the strengths\ of the first and second radio
frequency fields.

\section{Entanglement in a system of fictitious spins}

\subsection{Pure state}

The representation of a spin $\frac{3}{2}$ as two fictitious coupling spins $%
\frac{1}{2}$ allows us to investigate entanglement using the methods
developed for spin $\frac{1}{2}$ systems \cite{3.AmicoL2007,4.HorodeckiR2009}%
. The energy levels $E_{m}$ of Hamiltonian (3) are determined by a solution
of\ $\mathcal{H}$ $\left\vert \Psi _{m}\right\rangle =E_{m}\left\vert \Psi
_{m}\right\rangle $ ,\ where $\left\vert \Psi _{m}\right\rangle $\ are
eigenfunctions, $m=\frac{3}{2},\frac{1}{2},-\frac{1}{2},-\frac{3}{2}$. The
energy levels are

\begin{align}
E_{\pm\frac{3}{2}} & =\frac{1}{2}\left( -\omega_{0}\mp\sqrt{%
4\omega_{0}^{2}-6\omega_{0}\omega_{Q}\left( 2-\eta\right)
+\omega_{Q}^{2}\left( 12+\eta^{2}\right) }\right) ,\text{ \ \ }  \notag \\
E_{\pm\frac{1}{2}} & =\frac{1}{2}\left( \omega_{0}\mp\sqrt{%
4\omega_{0}^{2}+6\omega_{0}\omega_{Q}\left( 2-\eta\right)
+\omega_{Q}^{2}\left( 12+\eta^{2}\right) }\right)  \tag{11}
\end{align}
Thus, for a spin $\frac{3}{2}$ in an external field, $\omega_{0}\neq0$, the
states are non-degenerate. The energy level of $E_{+\frac{3}{2}}$
corresponds to the ground state described by the wave function

\begin{equation}
\left\vert \Phi_{\frac{3}{2}}\right\rangle =\left\vert 
\begin{array}{c}
a_{-} \\ 
b_{+} \\ 
b_{-} \\ 
a_{+}%
\end{array}
\right\rangle \text{,\ \ }  \tag{12}
\end{equation}
where%
\begin{equation*}
a_{\pm}=\mp\frac{1}{d}\sqrt{3}\left( \alpha-\eta\right) ,
\end{equation*}%
\begin{equation*}
b_{\pm}=\pm\frac{1}{d}\left( 6+\alpha+\sqrt{36+4\alpha\left( 3+\alpha
\right) -6\alpha\eta+3\eta^{2}}\right) ,
\end{equation*}%
\begin{equation*}
d=\sqrt{6\left( \alpha-\eta\right) ^{2}+2\left( 6+\alpha+\sqrt {%
36+4\alpha\left( 3+\alpha\right) -6\alpha\eta+3\eta^{2}}\right) ^{2}}.
\end{equation*}
$\allowbreak$Here $\alpha=\frac{\omega_{0}}{\omega_{Q}}$ is the normalized
external magnetic field.

To apply the methods of entanglement investigation, developed for spin-$%
\frac{1}{2}$ systems, we first map the Hilbert space for a spin $\frac{3}{2}$%
, which is four-dimensional, onto the Hilbert space for two fictitious spins 
$\frac{1}{2}$ according to Eqs. (8) - (10): $\left\vert \Phi _{\frac{3}{2}%
}\right\rangle =\left\vert \psi \right\rangle .$ The most general wave
function represented in terms of the standard basis for two fictitious spins
has the form \cite{29Valiev K.A}:%
\begin{equation}
\left\vert \psi \right\rangle =a_{-}\left\vert 00\right\rangle
+b_{+}\left\vert 01\right\rangle +b_{-}\left\vert 10\right\rangle
+a_{+}\left\vert 11\right\rangle  \tag{13}
\end{equation}%
\ \ \ \ \ \ \ \ \ \ \ \ \ \ \ \ \ \ \ \ \ \ \ \ \ \ \ 

\bigskip\ Concurrence $C$, a measure of entanglement, is determined by the
expression \cite{30.A. A. Kokin} 
\begin{equation}
C=2\left\vert a_{+}a_{-}-b_{+}b_{-}\right\vert .  \tag{14}
\end{equation}
\ 

Applying formulas (14) to wave function (12), we determine the concurrence
in the ground state of 3-dimensional Heisenberg model\ (10) which is
isomorphic to the ground state of a single nuclear spin $\frac{3}{2}$ in an
inhomogeneous magnetic and electric fields%
\begin{equation}
C=\frac{6+\alpha }{\sqrt{36+4\alpha \left( 3+\alpha \right) -6\alpha \eta
+3\eta ^{2}}}.  \tag{15}
\end{equation}%
Dependence of concurrence $C$ on an external magnetic field $\alpha $ and
asymmetry parameter $\eta $ is shown in Fig. 1. The concurrence decreases
with an increase of the external field and slowly depends on the asymmetry
parameter. At $\eta =0$ \ and low magnetic field ($\alpha <<1)$ \ the
maximum concurrence of 1$;$ with the increase of the asymmetry parameter the
maximum is shifted to higher magnetic field and it is observed at $\alpha
\simeq 1$ if $\eta =1.$ In a high magnetic field $\alpha >>1$ , \ $C=0.5$.
This differs from the results of the Ising and Heisenberg models \cite{23.M.
A. Yurishchev} which predict drop of the concurrence to zero at high fields.
The difference can be explained by the fact that, in the considered system,
the external field increases coupling of the fictitious spins and acts only
one spin (see (10)); the spins do not become completely aligned along the
field direction. However, Eq. (15) is not valid for strictly zero magnetic
field, $\omega _{0}=0,$ in which limit it gives the concurrence of $\left( 
\frac{\eta ^{2}}{12}+1\right) ^{-\frac{1}{2}}.$ At precisely $\omega _{0}=0$%
, pure NQR, no entanglement is present (the eigenstates are degenerate as
for the usual Ising Hamiltonian without any magnetic field, where
entanglement is absent \cite{24.D. Gunlycke}).\ Similarly to the Ising
model, the entanglement jumps from zero (at $\omega _{0}=0)$ to a finite
value even for an infinitesimal increase of a magnetic field, indicating the
quantum phase transition. Note, that when the external magnetic field is
applied along {\small direction }the{\small \ }$z$-axis, the entanglement
between spins is absent \cite{31.G. B. Furman,35.E. I. Kuznetsova}.

\subsection{Mixed state}

In real experiments temperature is finite, and a spin-$\frac{3}{2}$ system
is in a mixed quantum state. The system described by Hamiltonian (10) in the
thermodynamic equilibrium is characterized by the density matrix

\begin{equation}
\rho=Z^{-1}\exp\left( -\frac{\mathcal{H}_{H}}{k_{B}T}\right) ,  \tag{16}
\end{equation}
where $T$ is the spin temperature, $k_{B}$ is the Boltzmann constant, $Z=Tr%
\left[ \exp\left( -\frac{\mathcal{H}_{H}}{k_{B}T}\right) \right] $ is the
partition function.

To quantify the entanglement of the system of two fictitious spins, we will
use the concurrence $C_{T}$ defined by the following expression \cite{30.W.
K.Wootters}:%
\begin{equation}
C_{T}=\max \left\{ 0,\lambda _{1}-\sum_{j=2}^{4}\lambda _{j}\right\} 
\tag{17}
\end{equation}%
where $\lambda _{1}=\max \left\{ \lambda _{j}\right\} $ and $\lambda _{j}$ $%
\left( j=1,2,3,4\right) $ are the square roots of the eigenvalues of the
matrix 
\begin{equation}
R=\rho \left( \sigma _{y}\otimes \sigma _{y}\right) \bar{\rho}\left( \sigma
_{y}\otimes \sigma _{y}\right)  \tag{18}
\end{equation}%
where $\bar{\rho}$ \ is the complex conjugation of the density matrix (16).
\ 

Our calculation shows that the concurrence slowly depends on the asymmetry
parameter (Fig. 2). Fig. 3 \ shows dependences of the concurrence $C_{T}$ in
the mixed state on the normalized inverse temperature, $\beta =\frac{eQq_{ZZ}%
}{4I(2I-1)k_{B}T},$ and the normalized external magnetic field $\alpha $ at $%
\eta =0.14$. The system of fictitious spins is in the separable state
without applying external magnetic filed ($\omega _{0}$ = 0) at any
temperature. At applying sufficiently high magnetic field the entangled
state is observed; decrease in the field leads to a sudden disappearance of
the entangled state. The dependence of the critical inverse temperature $%
\beta _{C}$, which segregates the separable and entangled states, on the
magnetic field is presented in Fig. 4. At low magnetic fields $\alpha <0.1$
the critical temperature $T_{C}\symbol{126}1/\beta _{C}$ sharply decreases
with an increase of the field. At high magnetic fields $\alpha >>1$ the
dependence of the critical inverse temperature $\beta _{C}$ on the magnetic
field is well approximated by $\alpha \beta _{C}=0.85$. The concurrence
monotonically grows with an increase in $\beta $ above the critical value
(Fig. 5) while the dependence of the concurrence on magnetic field possesses
the maximum (Fig. 6). The maximum grows and moves to area of lower magnetic
field with an increase in $\beta $. At high fields $\alpha >>1,$ the
concurrence tends to $0.5$. The limit is independent of temperature and
equals to the concurrence limit for the system in the pure state at high
fields. \emph{\ }

\section{\textbf{Discussion and conclusions}}

According to the quantum mechanics rules, the mathematical description of a
system consisting of two particles is realized using the Kronecker product
of the operators of individual particles. We have explicitly shown that the
spin operators and the quadrupolar and Zeeman Hamiltonians for a spin $\frac{%
3}{2}$ can be represented in the operator terms for a system of two coupling
fictitious spins $\frac{1}{2}$ using the Kronecker product of the Pauli
matrices. Thus, the particle possessing spin $\frac{3}{2}$ can be considered
as consisting of two parts and we study the entanglement of these parts.

The representation of a single spin $\frac{3}{2}$ as a system of two
fictitious spins $\frac{1}{2}$ allowed us to study entanglement in the spin
system. One of the features of the fictitious spin system is that in both
the pure and the mixed states the concurrence tends to $0.5$ in a high
magnetic field $\alpha >>1.$

The calculation for $^{63}Cu$ in the five-coordinated copper ion site of $%
YBa_{2}Cu_{3}O_{7-\delta }$ at $\alpha =1$, $\eta =0.14$ and $eQq_{zz}=$ $%
62.8$ MHz \cite{32.Mali}, gives that the concurrence appears at $\beta =0.24$
(Fig. 5).$\,\allowbreak $This $\beta $ value corresponds to temperature $%
T\approx 2$ mK. It has been shown \cite{Fel'dman2007,31aFurman,33.G. B.
Furman,31.G. B. Furman,34G. B. Furman} that, for the XY and dipolar coupling
spin-$\frac{1}{2}$ systems entanglement appears at very low temperatures $%
T\sim 0.3\div 0.5$ 
%TCIMACRO{\U{b5}}%
%BeginExpansion
$\mu$%
%EndExpansion
K. This value is four orders smaller than the value estimated by us for a
quadrupole system.

The representation of a spin $\frac{3}{2}$ as a system of two coupling
fictitious spins $\frac{1}{2}$ and possibility of formation of the entangled
states in quadrupole systems open a way to the application of a single spin $%
\frac{3}{2}$ in quantum computation.

Caption figures

Fig. 1 The dependence of concurrence $C$ on the normalized external magnetic
field $\alpha$ and $\eta$ (pure state).

Fig. 2 Concurrence $C_{T}$ in the mixed state as a function of asymmetry
parameter $\eta$ at $\alpha$ =0.5 (a): $\beta$ =1 (solid black) ; $\beta$ =2
(green dashed); $\beta$ =3 (blue dotted); $\beta$ $=4$ (red dot-dashed) and
at $\beta$=2 (b): $\alpha$=0.5 (solid black) ; $\alpha$ =1 (green dashed); $%
\alpha$ =2 (blue dotted); $\alpha$ =3 (red dot-dashed).

Fig.3 The dependence of concurrence $C_{T}$ in the mixed state on plane $%
\beta$ and $\alpha$ at $\eta$=0.14.

Fig. 4 The phase diagram. The line presents boundary between the entangled
and separated states in the plane $\beta_{C}$ --$\alpha.$

Fig. 5 Concurrence $C_{T}$ in the mixed state as a function of inverse
temperature $\beta$ at $\eta$=0.14: $\alpha$=1 (solid black) ; $\alpha$ =2
(green dashed); $\alpha$ =3 (blue dotted); $\alpha$ =4 (red dot-dashed).

Fig. 6 Concurrence $C_{T}$ in the mixed state as a function of magnetic
field $\alpha $ at $\eta $=0.14: $\beta $ =1 (solid black); $\beta $ =2
(green dashed); $\beta $ =3 (blue dotted); $\beta $ =4 (red dot-dashed).

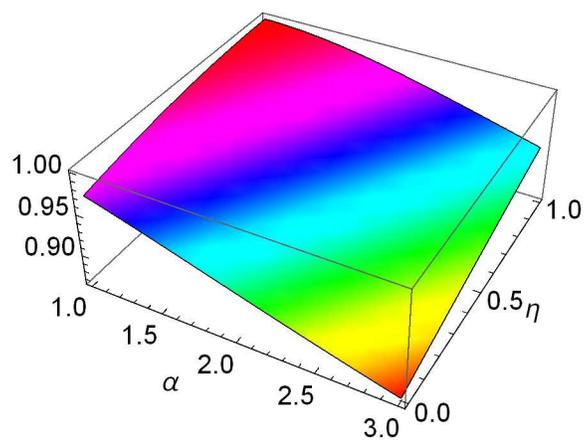
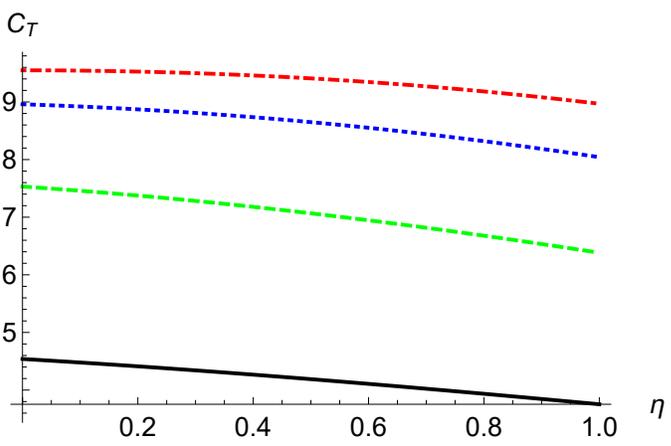
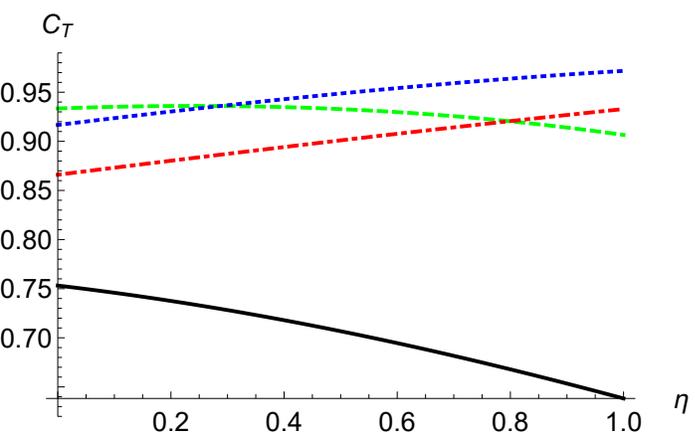
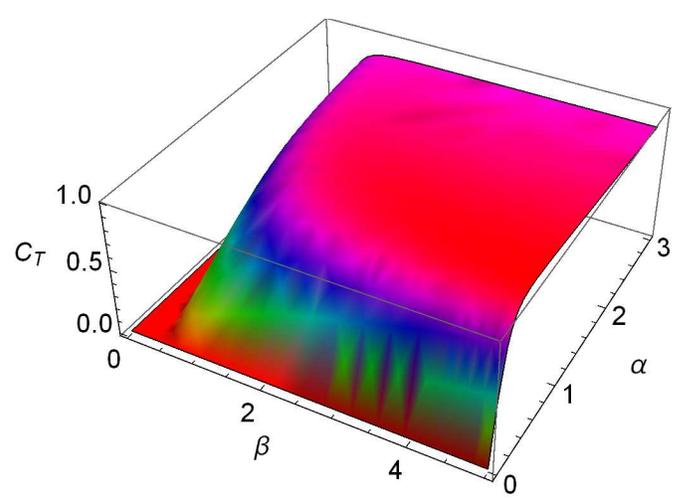
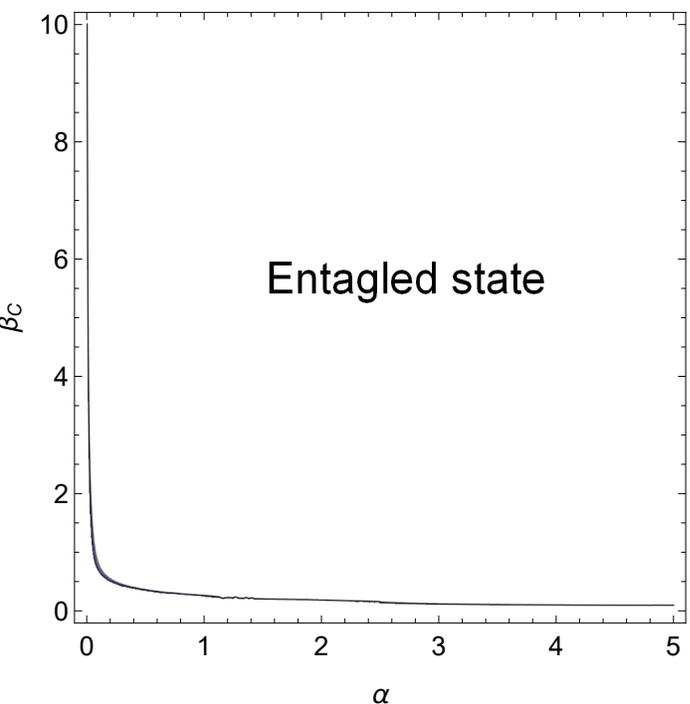
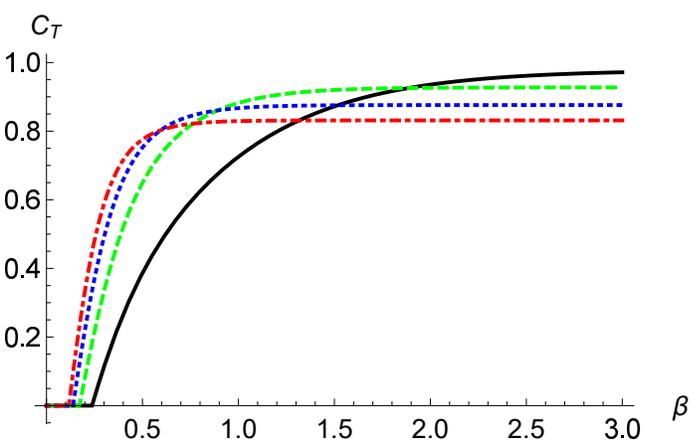
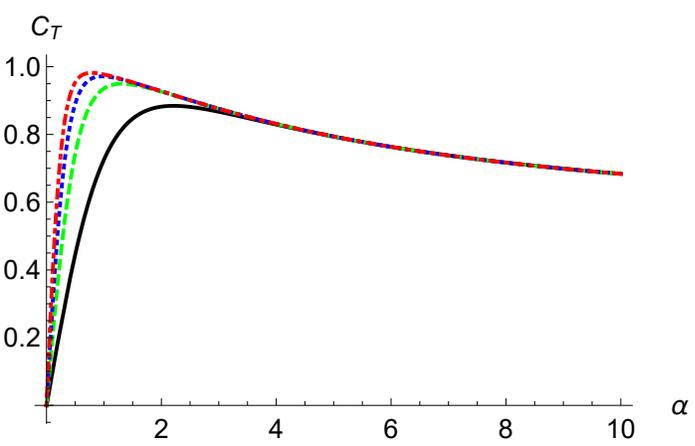
\end{document}